# Metamaterial-based octave-wide terahertz bandpass filters

Ali Maleki,[1] Avinash Singh,[1] Ahmed Jaber,[1] Wei Cui,[1] Yongbao Xin,[4] Brian T. Sullivan,[4] Robert W. Boyd,[1,2,3] Jean-Michel Ménard[1]

[1]Department of Physics, University of Ottawa, Ottawa, Canada
[2]School of Electrical Engineering and Computer Science, University of Ottawa, Ottawa, Canada
[3]Institute of Optics and Department of Physics and Astronomy, University of Rochester, Rochester, USA
[4]Iridian Spectral Technologies Ltd, 2700 Swansea Crescent, Ottawa, Canada

**Abstract**

We present octave-wide bandpass filters in the terahertz (THz) region based on bilayer-metamaterial (BLMM) structures. The passband region has a super-Gaussian shape with a maximum transmittance approaching 70% and a typical stopband rejection of 20 dB. The design is based on a metasurface consisting of a metallic square-hole array deposited on a transparent polymer, which is stacked on top of an identical metasurface with a sub-wavelength separation. The superimposed metasurface structures were designed using finite-difference time-domain (FDTD) simulations and fabricated using a photolithography process. Experimental characterization of these structures between 0.3 to 5.8 THz is performed with a time-domain THz spectroscopy system. Good agreement between experiment and simulation results is observed. We also demonstrate that two superimposed BLMM (2BLMM) devices increase the steepness of the roll-offs to more than 85 dB/octave and enable a superior stopband rejection approaching 40 dB while the maximum transmittance remains above 64%. This work paves the way toward new THz applications, including the detection of THz pulses centered at specific frequencies, and an enhanced time-resolved detection sensitivity towards molecular vibrations that are noise dominated by a strong, off-resonant, driving field.

**Introduction**

Periodical sub-wavelength structures, also referred to as one particular type of metamaterials, are deemed to play a crucial role for the advancement of terahertz (THz) technologies, notably in the fields of wireless communication [1,2], imaging [3], and signals processing [4]. These structures can be engineered to control several parameters of a THz pulse, such as its spectral amplitude, spatial phase, local field distribution, and polarization state [5,6]. Recently, metamaterials have been used to tune the properties of a THz pulse in both the spectral and temporal domains [7,8]. Other devices such as compact metalenses [9], chemical sensors [10], optical switching modulators [4], and perfect absorbers [11,12] were demonstrated when the metamaterial is confined to a single plane to form a metasurface. Metallic-based THz metasurfaces benefit from relatively low plasmonic loss in comparison to their counterparts operating in the visible or near-infrared regions. More importantly for many applications, these devices do not require precision nanofabrication tools since the metallic structures are in the range of tens of microns, scaling proportionally with the wavelength of operation, which is 300 µm (at 1 THz). The relatively large feature sizes of metasurfaces allows them to be fabricated with scalable and cost-effective metal deposition and photolithography techniques. Moreover, metasurface devices fill a special niche for spectral filtering applications in the THz region since multilayered dielectric filters, such as those commonly used in the visible and near-infrared regions, would require the deposition of millimeter-thick coatings, which is both costly and technically challenging. Previous works demonstrated that plasmonic modes in the THz region can be used to achieve tunable spectral filters [8], wide bandstop [13], and bandpass spectral filters [14]. Sharp bandpass spectral filters are especially important for signal detection since they can be used to select specific spectral components while rejecting unwanted frequencies that contribute to the noise. In standard time-resolved measurements in which all frequencies are effectively resolved simultaneously, this approach can significantly improve detection sensitivity over a given spectral region [15].

In recent years, multilayered plasmonic metasurfaces have been designed and fabricated to enable new coupled electric or magnetic modes that can contribute to create broadband resonances [16–18]. These filtering devices rely on two or more planar periodic arrays of sub-wavelength structures utilizing different shapes such as rectangles, crosses, or circular-slot structures [13,16,17,19]. They can also include phase-changing materials, such as vanadium dioxide, to achieve tunable spectral properties [20]. The performances of such filters are typically evaluated from three main criteria: (i) the maximum transmission power within a flat passband window, (ii) the rejection efficiency of frequency components in the stopband region, and (iii) the transition between the passband and stopband regions, also referred as the roll-off, which ultimately allows the device to distinguish desired spectral components from those nearby frequencies that must be blocked.  In this paper, the broadband THz filtering properties of a bilayer metamaterial (BLMM) device is demonstrated featuring ~70% flat-top transmittance over an octave-spanning bandwidth. The stopband attenuation is greater than 10 dB over the spectral window of interest (0.3 to 5.8 THz) with some spectral regions featuring more than 20 dB attenuation. This device relies on two identical plasmonic metasurfaces, both based on a square-loop hole periodic array, separated by a transparent polymer ('coupling') layer. The BLMM design is first optimized (using numerical FDTD simulations) for broadband transmission in the THz region; fabricated with standard photolithography techniques; and, finally, characterized utilizing a time-domain THz system. An excellent agreement is observed between theoretical and experimental results. In addition, it is demonstrated that stacking two BLMMs dramatically improves the rejection efficiency in the stopband region while the maximum transmittance remains relatively unaffected.

**Experiment**

Figure 1(a) shows a schematic of the BLMM broadband filters, which consist of two identical plasmonic metasurfaces separated by a thin dielectric coupling layer. A photolithography process utilizing a negative lift-off resist to pattern the desired metasurface (square-loop hole arrays) on a 188 µm thick Zeonor substrate (Zeonor is a cyclo-olefin copolymer transparent to THz radiation with a refractive index of n ~ 1.53 at 1 THz). Aluminum is next deposited onto the patterned substrate using a sputtering process. This is then followed by a lift-off process which results in the desired metallic square-loop hole arrays which have a 4-fold symmetry that ensures a linear optical response independent of the incident polarization state. Figure 1(b) shows the topography of the fabricated metasurface as measured with an atomic force microscope (AFM). Depth profile analysis of the structures, as shown in Fig. 1(c), indicates a uniform aluminum thickness of ~220 nm.

In order to create a steeper bandpass fall-off, a 'double-cavity' (BLMM) structure is created by utilizing two identical metasurfaces (described above) in which the metasurface layers are separated by a thin dielectric 'coupling' layer.  The 'coupling' layer in this paper is based on a thin (~10-33 µm) double-sided tape that binds two metasurfaces together. Previous work has demonstrated that the spectral properties of such a bi-layered structure is relatively robust to misalignment between the structural elements of the stacked metasurfaces [19,21]. Nonetheless, a mask aligner was used to achieve a sub-micron positioning of the overlapping square-loop hole pattern to ensure good reproducibility between different samples and optimal agreement between experimental measurements and numerical simulations.

The BLMM devices are characterized utilizing time-domain THz spectroscopy. The experimental setup relies on an optical source delivering 60 fs pulses centered at a wavelength of 1030 nm [22,23]. Optical rectification in a 200 µm-thick 110-orientated GaP crystal generates broadband THz transients detected by electro-optic sampling relying on an identical nonlinear detection crystal. Numerical simulations are performed with a three-dimensional (3D) finite-difference time-domain (FDTD) solver from Lumerical Inc. to investigate and optimize the target design parameters and achieve the desired broadband filtering properties in the THz region. As shown in the inset of Fig. 1a, the geometry of the metallic structure is defined from its lattice pitch ($P$), length of the outer square-loop hole ($L$), length of the inner square-loop hole ($w$), and the thickness of the dielectric coupling layer ($d$) separating the two metasurfaces. Periodic boundary conditions along the in-plane axes are employed in the simulations as well as a perfectly matched layer (PML), which is applied in the direction of optical propagation to absorb all incident fields and prevent reflection at that interface.

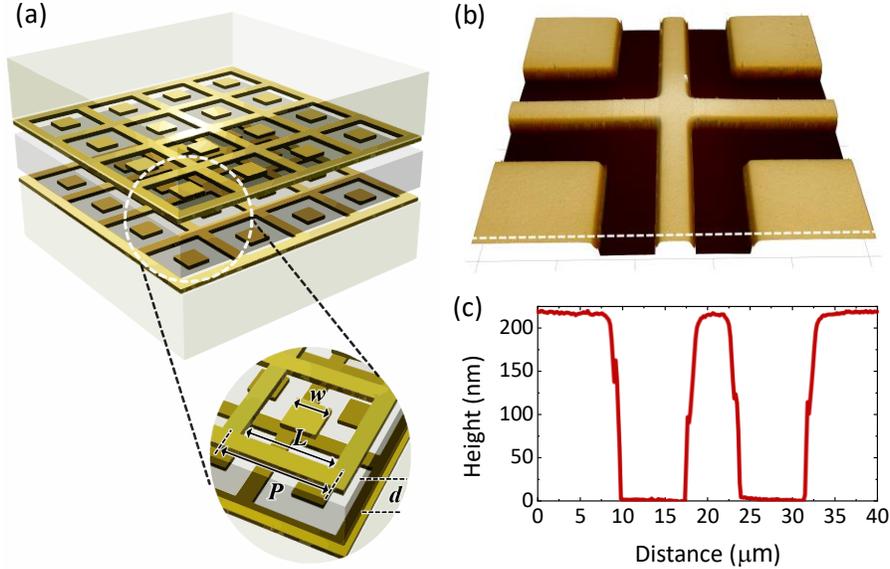

Fig. 1. (a) General schematic of the bilayer-metamaterial (BLMM) bandpass filter. Two plasmonic metasurfaces are placed adjacent to each other separated by a dielectric coupling layer of thickness *d*. Each metasurface consists of an array of metallic square-loop hole structure deposited on Zeonor, a THz transparent substrate. The lattice pitch (*P*), length of the outer square-loop hole (*L*), and inner square-loop hole (*w*) of the array are labeled with black arrows. (b) Three-dimensional (3D) atomic force microscope (AFM) image of sample S2 and (c) corresponding depth profile indicating a metal depth of ~220 nm.

**Results and discussion**

The metasurface geometry used for the fabrication of six BLMM devices, labelled S1 to S6, is presented in Table 1. Figures 2(a) and (b) show that the measured THz transmission spectrum (circles) corresponding to each structure is in good agreement with FDTD simulation results (solid lines). We find that the ratio between the highest and lowest frequencies at half the maximum of the transmittance correspond to 2, 2.2, 1.9, 1.7, 1.9, 2.8 for the devices S1 to S6, respectively. Therefore, the BLMM designs demonstrate a FWHM bandwidth that corresponds to approximately one octave (or a factor of 2). The transmittance spectra exhibit a super-Gaussian shape with a measured transmittance >70% in the passband region. This relatively high transmittance is partially attributable to the weak Fresnel reflection at the air-polymer interface, which is significantly lower than that typically observed at an air-semiconductor interface [8]. Additionally, the BLMM devices exhibit steep roll-offs (as measured from the high-frequency side of the passband) of >50 dB/Octave and up to 100 dB/Octave for S6. Such roll-off values are larger than those reported for other double-layered metamaterial devices based on structures such as skewed circular slot rings (30.2 dB/Octave), meandered slots (44.6 dB/Octave), and Jerusalem cross slots (58.3 dB/Octave) [19]. These roll-offs are also comparable to those reported for more complex multiple-layered metamaterial structures [24]. As shown in Fig 2(b) for S1, S2 and S3, such a roll-off leads to a significant transmittance suppression, approaching 30 dB within an octave of the central transmitted frequency. Furthermore, the attenuation of these devices remains relatively high (>10 dB) at frequencies extending beyond an octave from the passband region. Figure 2(c) shows the measured maximum transmittance ($T_{max}$) of the six devices as a function of frequency, in which circles indicate the center frequency of the passband filters and horizontal lines correspond to the FWHM linewidth of the devices. Another figure of merit used to characterize bandpass filters is the fractional bandwidth (FBW) defined as the absolute bandwidth at FWHM divided by the central frequency. As shown in Fig. 2(d), our structures feature a FBW > 50%, and up to 93% for S6, which is, to our knowledge, the largest FBW reported for a THz bandpass filter. In comparison to single-layer metasurface structures, the BLMM design provides a broad bandpass region with higher attenuation in the stop-band region, a flatter passband and steeper roll-offs [25–27].

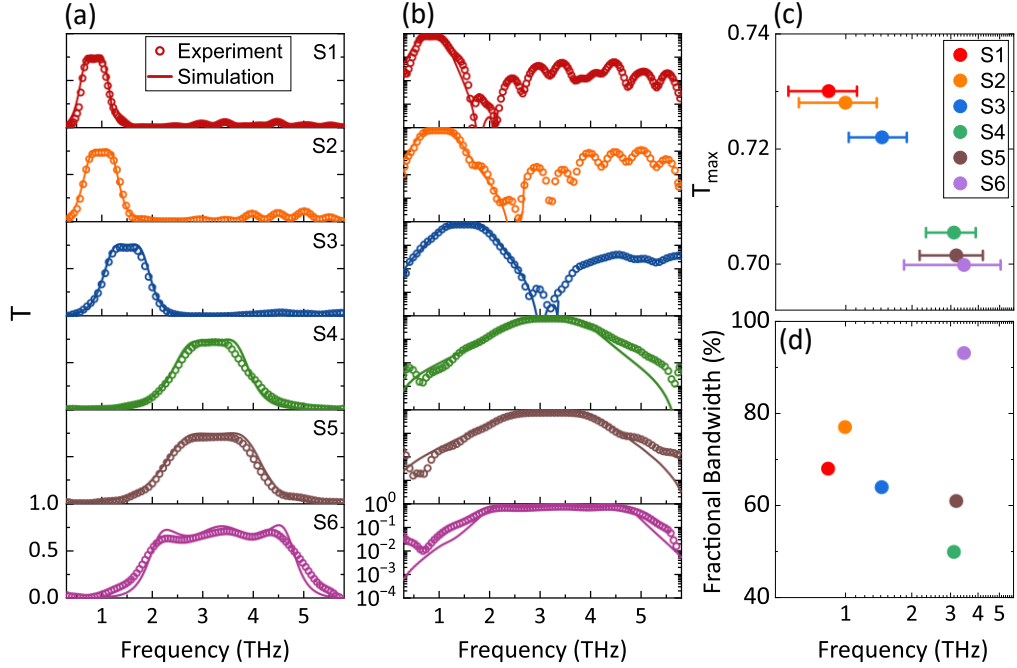

Fig. 2. THz measured (circles) and simulated (solid lines) transmittance spectrum of the BLMM-based broad bandpass filters in (a) linear scale and (b) semi-logarithm scale. Experiments are performed with a time-domain THz spectroscopy system and are in good agreement with FDTD simulation results. (c) Measured maximum transmission ($T_{max}$) of the structures over center frequency in semilogarithmic scale, in which error bars present the corresponding octave-spanning FWHM linewidth. (d) Calculated fractional bandwidth (in percentage) of all BLMM devices over frequency for the experiment results.

Table 1. Geometrical parameters of the broad bandpass devices where $P$ is the lattice pitch, $L$ is the outer square-slot length, $w$ is the inner square-slot length, and $d$ is the distance between two metasurfaces as shown schematically in Fig. 1(a).

| Device | $P$ (μm) | $L$ (μm) | $w$ (μm) | $d$ (μm) |
|---|---|---|---|---|
| S1 | 86 | 80 | 60 | 33 |
| S2 | 70 | 66 | 48 | 33 |
| S3 | 52 | 48 | 36 | 23 |
| S4 | 29 | 23 | 17 | 10 |
| S5 | 28 | 24 | 16 | 10 |
| S6 | 32 | 28 | 15 | 30 |

The geometrical dimensions shown in Table 1 were determined using numerical simulations to obtain a broad, sharp and symmetrical bandpass region centered at different frequencies. Due to scale invariance, decreasing the size of all geometrical parameters by the same amount proportionally shifts the device's transmittance profile to lower wavelengths (higher frequencies). The structures S1, S2 and S3 are based on a common metasurface design displaying similar ratios between geometrical parameters. As a result, these three filters display similar properties in terms of bandwidth, roll-offs, and overall transmittance profile. Limitations in the wafer-size photolithography process prevented using the same filter design at higher frequencies since defects and spatial homogeneity across the devices play a larger role as the structure parameters are decreased. More specifically, it was established that the minimum feature sizes of (*P-L*) should not be less than 4 µm. The devices S4, S5, and S6 are therefore designed utilizing a different model to explore filtering properties centered around 3 THz, while ensuring the minimum feature is large enough to ensure the fabrication of high-quality samples. Note also that the coupling parameter *d* cannot be fully controlled as it based on the thickness of the adhesive region between the two metasurfaces.

A 2D map of the electric field spatial distribution and surface current distribution on the metallic layer of the structure S5 is shown in Fig. 3. Focusing on the spectral components within the bandpass region of the filter, Fig. 3(a) shows a top view of the device with localized surface resonances at 2.9 THz. A field enhancement of 26 (relative to the incoming field amplitude) is observed at the corners of the central square structure along with an anisotropic field distribution with higher field amplitudes along the lateral sides of the structures due to the incident horizontal polarization. Electrical currents are induced at the outer edge of the central square and the inner edge of the surrounding frame with similar magnitude but opposite directions. Figures 3(b) and (c) show cross section of the electric field amplitude and direction at frequencies of 2.9 and 3.6 THz, respectively, corresponding to the boundaries of the filter passband region. This side view is taken at the edge of the inner square structures with a 3D schematic of the unit cell (in grey) superimposed. Spectral filters relying on stacked metasurfaces are able to support effective cavity modes dominated by electric dipole resonances, magnetic dipole resonances, and standing-wave-like resonances, as previously discussed in the literature [13,16,17,19,28,29]. In Fig. 3(b), the symmetric oscillation of the induced current in the top and bottom layers indicate an electric dipole resonance. At a slightly up-shifted frequency, Fig. 3(c) shows an anti-symmetric resonance where electrical currents in the top and bottom layers oscillate in opposite directions. Such mode is related to a magnetic dipolar resonance. It is the interference between these two modes, also referred to as trapped modes [29,30], that lead to such broad and sharp passband region centered at 3.1 THz. Note also that the surrounding frame creates a standing-wave-like resonance, which increases the absorption in the stopband region [16,19].

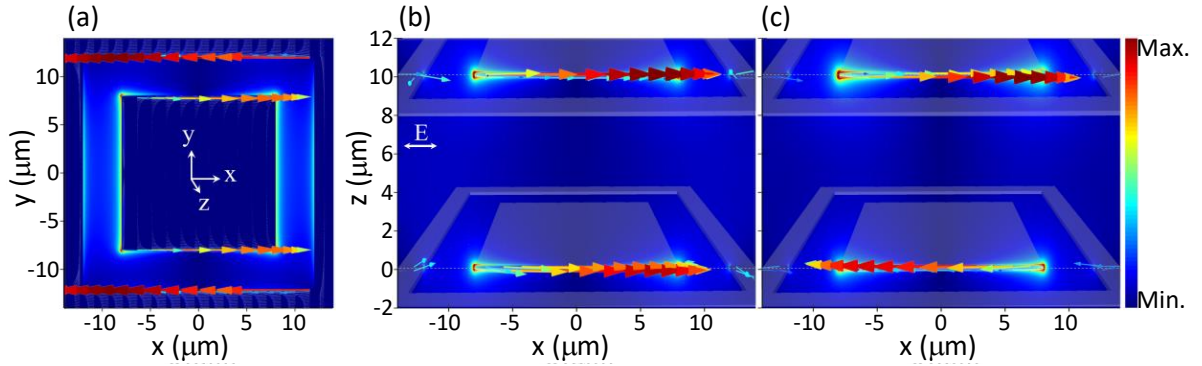

Fig. 3. Normalized electric field amplitude distribution of the BLMM structure S5. The arrows indicate surface current distribution within one period of the array. (a) Top view distribution in xz-plane at the surface of the bottom layer plasmonic structure (*z=0.2* µm) at the frequency of 2.9 THz. (b) and (c) side view of the electrical field distribution and direction in xz-plane, corresponding to a 2D cross section taken at *y=-8* µm at a frequency of 2.9 THz and 3.6 THz, respectively. The grey 3D schematics in (b) and (c) illustrate the simulated unit cell of the stacked metasurfaces. The incident THz wave propagates in the z-direction and is polarized along the x-direction.

To create devices with improved spectral filtering performances, two BLMMs were joined together with a thin adhesive film (3TC-J0005TTG) to fabricate a '2BLMM' device shown schematically in Fig. 4(a). Based on the thickness of the Zeonor substrate and the adhesive film, the two BLMM structures are separated by a distance *D=386* µm, which is large enough to prevent near-field effects. The configuration is therefore independent of the relative cross structure alignment and orientation. Figure 4 compares the THz transmittance measured with S2 and S4 as individual BLMM devices (blue curves) and the transmission measured in a 2BLMM configuration (orange curves). The 2BLMM devices show improved filtering performances in terms of background attenuation. For example, the stopband region provides stronger attenuation of the signal by approximately two orders of magnitude

(dotted lines in Figs. 4b and c). Also, the roll-offs (as measured from the high-frequency side) now correspond to 86 dB/Octave and 85 dB/Octave for S2-2BLMM and S4-2BLMM, respectively, an improvement corresponding roughly to the square value of the roll-offs measured with the corresponding BLMM devices. This stronger stopband attenuation and the sharper transitions between the passband and stopband regions is consistent with a design based on two independent BLMMs, where the transmittance spectrum then corresponds to the square transmittance of a single device. There are, however, significant advantages in stacking and binding the two BLMMs as shown in Fig. 4. First, the maximum transmittance in the passband region is kept relatively high (>65%) as there is no air gap thus avoiding additional Fresnel reflections. Also, a stronger than expected attenuation is observed in the stopband region: for S2 the transmittance between 4 and 5.5 THz is reduced from $10^{-1}$ to $<5 \times 10^{-4}$ (or 10 dB to 33 dB) when using a 2BLMM geometry, which exceeds the expected signal reduction based on a square dependence. This enhanced signal reduction in the stopband region may be indicative of a dipole interaction between the two BLMM devices inducing a new coupled mode that contributes to reduce transmittance across this spectral region.

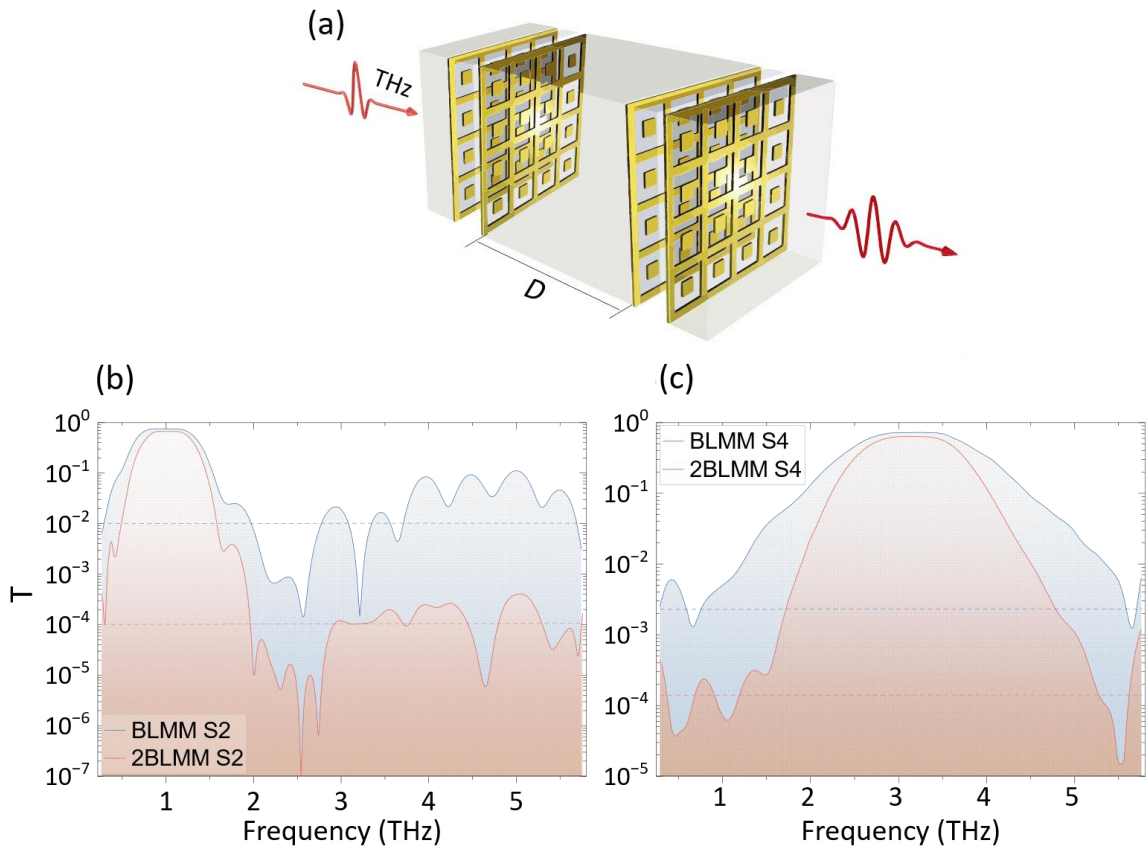

Fig. 4. (a) Schematic of the 2BLMM devices fabricated from two BLMM structures bound together with double-sided tape (separated by a length '$D$'). Comparison of the measured THz transmission of the BLMM and 2BLMM devices for (a) S2 (blue curve) and (b) S4 (orange curve). Dotted lines show the averaged attenuation floor.

**Conclusion**

We have designed and experimentally demonstrated six bilayer metamaterial (BLMM) devices with THz bandpass filtering properties. These devices feature a flat passband region centered at different frequencies, with a maximum transmittance exceeding 70% and a large spectral bandwidth corresponding to approximately one octave. The transmittance spectrum has a super-Gaussian profile featuring steep roll-offs, while the stopband region offers a strong signal suppression (generally > 20 dB) and likely extends much beyond the investigated spectral region of

0.3 to 5.8 THz. Experimental measurements and FDTD simulations are in good agreement. The fabrication of the BLMM devices is relatively straightforward as it relies on a standard photolithography process. Furthermore, it has been demonstrated that two BLMM devices ('2BLMM') can be joined together to improve the spectral filtering selectivity. This 2BLMM configuration significantly enhances the signal suppression in the stopband region while the transmission remains flat and high (>65%). We believe this type of filter design will play a major role in selecting THz signals contained within a known spectral region to notably enable sensitive wireless communication systems [2]. In time-domain spectroscopy, spectral filtering can also lead to significant improvement of the signal-to-noise by notably blocking spectral components necessary to drive a nonlinear system while investigating the signal at higher harmonics [15].

**Funding, Acknowledgments, and Disclosures**

We are thankful to Ksenia Dolgaleva and Orad Reshef for insightful discussions. The authors acknowledge support from the Natural Sciences and Engineering Research Council of Canada (NSERC) Strategic Partnership Program (STPGP/ 521619-2018). J.-M.M. acknowledges funding from the NSERC Discovery funding program (RGPIN-2016-04797) and Canada Foundation for Innovation (CFI) (Project Number 35269). A.M acknowledges support of the Ontario Graduate Scholarship (OGS), the University of Ottawa Excellence Scholarship, and the University of Ottawa International Experience Scholarship. The authors declare no conflicts of interest.